\documentclass[epj,twocolumn]{webofc}
\usepackage[varg]{txfonts}   
\woctitle{XLIV International Symposium on Multiparticle Dynamics}
\begin{document}
\title{Multiparticle production in nuclear collisions using 
effective-energy 
\\approach}
%
%

\author{Aditya Nath Mishra\inst{1}\fnsep\thanks{\email{Aditya.Nath.Mishra@cern.ch}} \and
        Raghunath Sahoo\inst{1}\fnsep\thanks{\email{Raghunath.Sahoo@cern.ch}} \and
        Edward K.G. Sarkisyan\inst{2,3}\fnsep\thanks{\email{sedward@mail.cern.ch}}
    \and Alexander S. Sakharov\inst{2,4}\fnsep\thanks{\email{Alexandre.Sakharov@cern.ch}}
}
\institute{Discipline of Physics, School of Basic Sciences, Indian Institute of Technology Indore, Indore-452017, India
\and  Department of Physics, CERN, 1211 Geneva 23, Switzerland
\and  Department of Physics, The University of Texas at Arlington, Arlington, TX 76019, USA 
\and  Department of Physics, Kyungpook National University, Daegu 702-701, 
Korea }

\abstract{%
 The dependencies of charged particle pseudorapidity density and transverse energy 
pseudorapidity density at midrapidity on the collision energy and on the number of nucleon 
participants, or centrality, measured in nucleus-nucleus collisions are studied in the energy 
range 
 spanning a few GeV to a few TeV per nucleon. 
 The study
is based on the earlier proposed 
 model,
 combining the constituent quark picture 
together with Landau relativistic hydrodynamics and shown to interrelate 
the measurements from 
different types of collisions. Within this picture, the dependence on the number of participants 
in  heavy-ion collisions  are found to be well described in terms of the effective energy defined 
as a centrality-dependent fraction of the collision energy. 
 The 
effective-energy approach  
 is shown to reveal 
 a similarity in the energy dependence  
 for the most 
 central 
 and centrality data in the entire available energy range. Predictions are 
 made 
 for the forthcoming higher-energy measurements in heavy-ion 
collisions at the LHC.
}
\maketitle

\noindent
{\textbf 1.} High-energy multiparticle production
 attracts considerable 
interest, as, on the one hand, the 
 bulk 
observables measured
 there 
 bring 
important information on the underlying dynamics of strong interactions, while on the other hand,  
this process still eludes its complete understanding. It is already more than half a century as 
the multiplicity of the produced particles are considered to be derived by the collision energy
\cite{Heisenberg,Fermi,Landau}.  In this picture the energy pumped into the collision zone in the 
very first stage of the collision defines the volume of the interaction lump of participant 
patterns. Later on, the approach of ``wounded'' nucleons, or nucleon participants,  has been 
proposed to describe the multiplicity and particle distributions \cite{wound}. In this approach the 
multiplicity is expected to be proportional to the number of participants. 
However, it was 
observed at RHIC and similarly at LHC energies, the concept of wounded nucleons does not describe 
the measurements where the data found to demonstrate an increase with the number of nucleon 
participants. Multiparticle production in nucleus-nucleus collision, hence, cannot be explained as a mere
superposition of proton-proton collisions. The problem has been addressed in the nuclear overlap model using Monte Carlo 
simulation in the constituent quark framework, and the scaling has been shown to be restored 
\cite{eremin,ind,indpart,nouicer}. In addition, it was observed that the multiplicity and 
midrapidity-density distributions are similar in $e^+e^-$ and in the most central (head-on) 
nuclear collisions \cite{phobos-sim} at the same center-of-mass (c.m.) energy pointing to the 
universality of multihadron production. However, the expectation to 
observe this type of 
universality in hadronic and nuclear collisions at similar c.m. energy per nucleon has not been 
shown by the data where the measurements in hadron-hadron collisions show 
significantly lower values compared to those in central heavy-ion
 collisions \cite{book,advrev}.     

To interpret these observations, the energy dissipation approach of constituent quark 
participants has been proposed in \cite{edward} by two of the authors of 
 this 
 report. In this 
picture, the process of particle production is driven by the amount of energy deposited by 
interacting participants into the small Lorentz-contracted volume during the early stage of the 
collision. The whole process of a collision is then considered as the expansion and the 
subsequent break-up into particles from an initial state. This approach resembles the Landau  
phenomenological hydrodynamic approach  of multiparticle production in relativistic particle 
interactions \cite{Landau}.
 In the picture proposed in \cite{edward}, 
the Landau hydrodynamics  is combined with the constituent quark model \cite{constitq}. This makes  
the secondary particle  production  to be basically driven by the amount of the initial {\it 
effective} energy deposited by participants -- quarks or nucleons, into the Lorentz contracted 
overlap region. In ${pp/\bar{p}p}$ collisions, a single constituent (or dressed) quark from each 
nucleon takes part in a collision and rest are considered as spectators. Thus, the effective 
energy for the production of secondary particles is the energy carried by a single quark pair 
{\sl i.e.} 1/3 of the entire nucleon energy. 
Contrary, in the head-on heavy-ion collisions, the participating nucleons are considered 
colliding by all three constituent quarks from each nucleon which makes the whole energy of the 
colliding nucleons (participants) available for secondary particle production. Thus, one expects 
that bulk observables measured in the head-on heavy-ion collisions at the c.m. energy per 
nucleon, $\sqrt{s_{NN}}$, to be similar to those from  ${pp/\bar{p}p}$ collisions but at a three times 
larger c.m. energy {\sl i.e.} $\sqrt{s_{pp}} \simeq 3\sqrt{s_{NN}}$.

Combining the above-discussed 
ingredients of the constituent quarks and  Landau hydrodynamics, one obtains the relationship 
between charged particle rapidity density per participant pair, $\rho(\eta)=(2/
N_{\rm{part}})dN_{\rm{ch}}/d\eta$ at midrapidity ($\it{\eta} \approx $~0) in heavy-ion collisions 
and that in ${pp/\bar{p}p}$ collisions:
 \begin{equation}
 \frac{\rho(0)}{\rho_{pp}(0)} = 
 \frac{2N_{\rm{ch}}}{N_{\rm{part}}\, N_{\rm{ch}}^{pp}} 
 \, \sqrt{\frac{L_{pp}}{L_{NN}}} \, .
 \label{eqn1}
 \end{equation}
In Eq.(\ref{eqn1}) the relation of the pseudorapidity density and the mean multiplicity is 
applied in its Gaussian form as obtained in Landau hydrodynamics. The factor $L$ is defined as 
$L = {\ln}({\sqrt{s}}/{2m})$. According to the approach considered, $m$ is the proton mass, 
$m_{p}$, in nucleus-nucleus collisions and the constituent quark mass in ${pp/\bar{p}p}$ collisions 
which is set to $\frac{1}{3}m_{{p}}$. $N_{\rm{ch}}$ and $N_{\rm{ch}}^{pp}$ are the mean 
multiplicities in nucleus-nucleus and nucleon-nucleon collisions, respectively,  and 
$N_{\rm {part}}$ is the number of participants. Then
 one evolves Eq.~(\ref{eqn1}) for the 
rapidity density $\rho(0)$ and the multiplicity $N_{\rm{ch}}$ at $\sqrt{s_{NN}}$, and the 
rapidity density $\rho_{pp}(0)$ and the multiplicity
$N_{\rm{ch}}^{pp}$ at $3 \sqrt{s_{NN}}$: 

\begin{eqnarray}
\nonumber
&\rho(0)& = \rho_{pp}(0)\, 
\frac{2N_{\rm{ch}}}{N_{\rm{part}}\, N_{\rm{ch}}^{pp}} 
\, \sqrt{1 - \frac{4 \ln 3}{\ln (4m_p^{2}/s_{NN})}}\,,  \\
    & & \sqrt{s_{NN}}=\sqrt{s_{pp}}/3 \, .
\label{eqn3}
\end{eqnarray} 

It was found in Ref. \cite{edward} that the current approach is able to reproduce
very well the data on the c.m. energy dependence of the midrapidity density
 measured in the most central heavy-ion collisions by interrelating by 
Eq.~(\ref{eqn3}) the measurements in hadronic and nuclear collisions up 
to the top RHIC energy. Moreover, it was also shown that similarly, the 
total  multiplicities in these types of collisions follow the energy-dependence 
 universality. 
  Such a universality is found to  correctly predict \cite{edward} 
the value of the midrapidity density in {\it pp} interactions at the
 TeV LHC energies \cite{cms-conf}. 
 
In this report, 
 we extend the above-discussed approach of the
constituent quark participants and Landau hydrodynamics 
to the midrapidity pseudorapidity-density dependence on the number of 
(nucleon) participants. Based on this energy dissipation picture,
we apply effective-energy consideration to the pseudorapidity 
density of the transverse energy at midrapidity.
 We give predictions for 
 foreseen 
 higher-energy heavy-ion collisions
 at the LHC
considered here 
at 
$\sqrt{s_{NN}}= $~5.13 TeV, corresponding to the scheduled 13 TeV 
{\it pp} 
collision LHC restart. The predictions for heavy-ion collisions at 
$\sqrt{s_{NN}}=$~5.52 TeV, corresponding to 7 TeV $p$-beam, are made
elsewhere \cite{edward14}.      
\\

\begin{figure*}
\centering
\includegraphics[width=10cm]{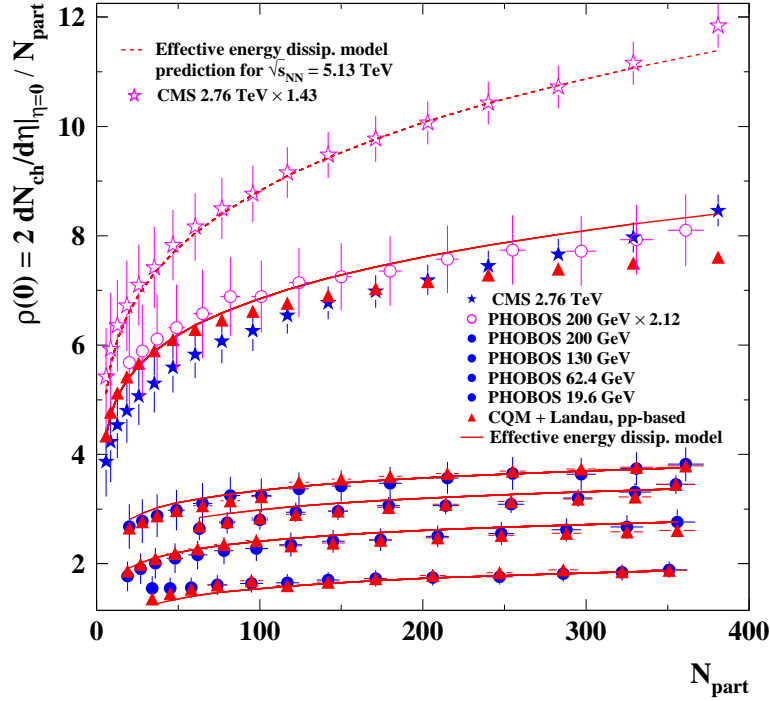}
\caption{ The charged particle pseudorapidity density at midrapidity per participant pair as a 
function of the number of participants, $N_{\rm part}$. The solid circles show the dependence 
measured in  AuAu collisions at RHIC by PHOBOS at $\sqrt{s_{NN}}=19.6$ to 200~GeV \cite{phobos-all} (bottom 
to top). The solid stars show the measurements in PbPb collisions at LHC by CMS at 
$\sqrt{s_{NN}}=2.76$~TeV  \cite{cms276c}. The solid triangles show the calculations by Eq.~(\ref{eqn4}) 
using ${pp/\bar{p}p}$  data. The lines represent the effective-energy 
dissipation approach predictions    
based on the hybrid fit to the c.m. energy dependence of the midrapidity density in central 
heavy-ion collisions shown in Fig.~\ref{Fig2}. The open circles show the PHOBOS measurements at 
$\sqrt{s_{NN}}=200$~GeV multiplied by 2.12, while the open stars show the CMS measurements multiplied by 
1.43.}
\label{Fig1}       
\end{figure*}

\begin{figure*}
\centering
\includegraphics[width=10cm]{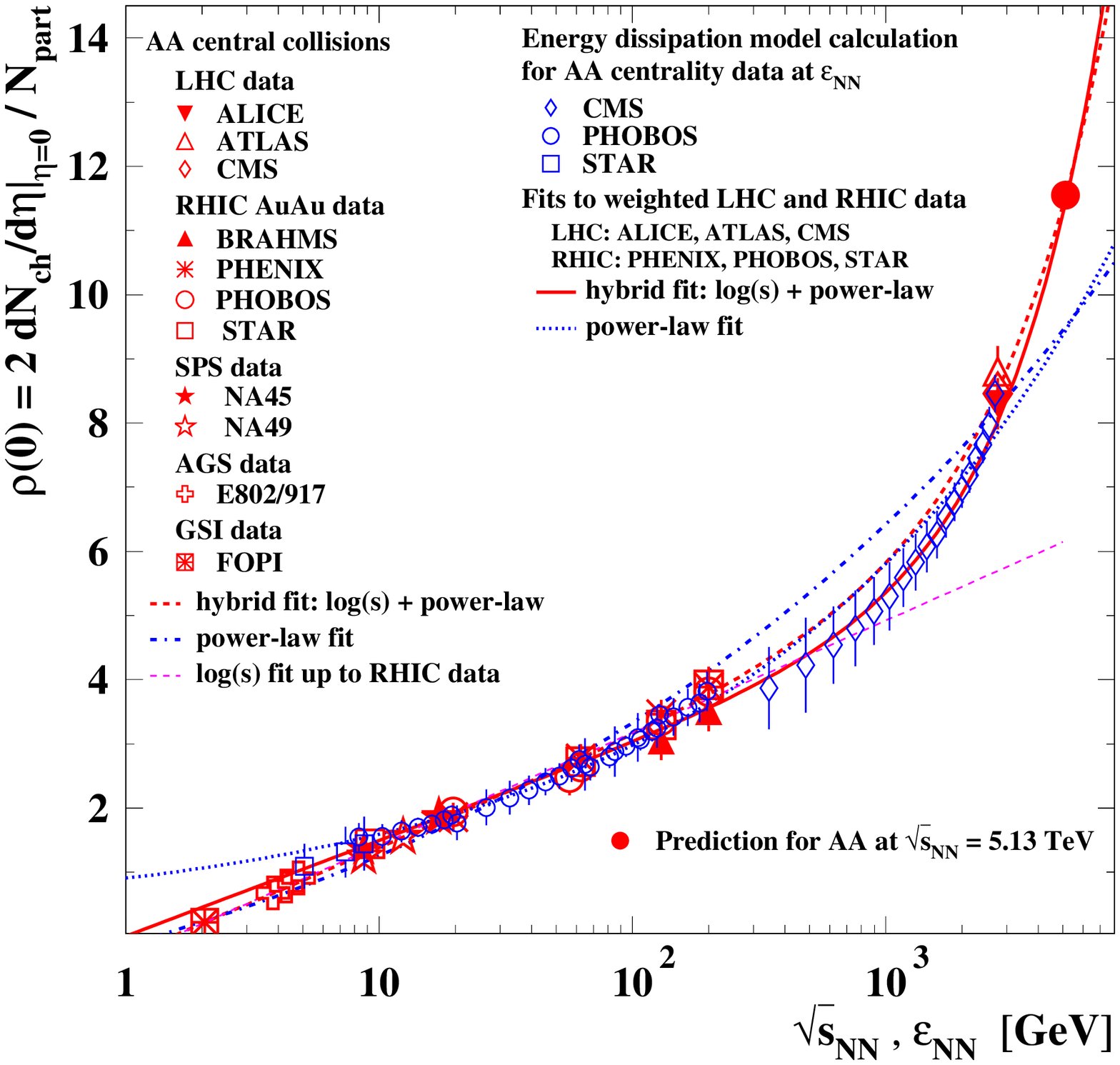}
\caption{The charged particle pseudorapidity density per participant pair at midrapidity as a 
function of c.m. energy per nucleon, $\sqrt{s_{NN}}$, in central nucleus-nucleus (AA) collisions (shown 
by large symbols), and as a function of effective c.m. energy, $\varepsilon_{NN}$ (Eq.~(\ref{Eeff})), for AA 
collisions at different centrality (small symbols). The data of central AA collisions  are from: 
the PbPb measurements at LHC by ALICE \cite{alice276r}, ATLAS \cite{atlas276c}, and CMS \cite{cms276c}
experiments; the AuAu measurements at RHIC by BRAHMS \cite{brahms}, PHENIX \cite{phenix-all}, PHOBOS 
\cite{phobos-all}, and STAR \cite{star-all,star9.2} experiments; the values recalculated in 
\cite{phenix-all} from the measurements at CERN SPS by CERES/NA45 \cite{na45} and NA49 \cite{na49} 
experiments, at Fermilab AGS by E802 and E917 experiments \cite{ags}, and at GSI by FOPI Collab. 
\cite{fopi}.
The centrality data represent the measurements by CMS \cite{cms276c}, PHOBOS \cite{phobos-all}, and 
STAR \cite{star9.2}; the CMS and PHOBOS data are those from Fig.~\ref{Fig1}, while for clarity,just 
every second point of the PHOBOS measurements  is shown. The dashed-dotted line and the dashed 
line show the fits to the central collision data: the power-law fit, $\rho(0) =-2.955 +2.823\, 
s_{NN}^{0.087}$, and the hybrid fit, $\rho(0)=-0.306+ 0.364\ln(s_{NN}) +0.0011\,  s_{NN}^{0.5}$. The 
thin dashed line shows the linear-log fit, $\rho(0)=-0.327 +0.381\, 
\ln(s_{NN})$ \cite{edward} to 
the central collision data up to the top RHIC energy. The dotted line and the solid line show the 
fits to the centrality data: the power-law fit, $\rho(0)=0.244 +0.663\, \varepsilon_{NN}^{0.308}$, and the 
hybrid fit, $\rho(0)=0.002+ 0.646 \ln(\varepsilon_{NN}) +0.0003 \, \varepsilon_{NN}^{1.158}$, respectively. The fitted 
centrality data include, except of the shown data,  also the measurements by ALICE \cite{alice276c} 
and ATLAS \cite{atlas276c}  at the LHC, and by PHENIX \cite{phenix-all} and STAR \cite{star-all, 
star9.2} at RHIC (not shown). The solid circle shows the prediction for 
$\sqrt{s_{NN}}=5.13$~TeV.}
\label{Fig2}       
\end{figure*}

\noindent
{\textbf 2.} In Fig.~\ref{Fig1}, the charged particle pseudorapidity 
density per participant pair at 
midrapidity as a function of number of participants is shown as   measured by PHOBOS experiment 
in AuAu collisions at RHIC at c.m. energy of $\sqrt{s_{NN}}=19.6$ to 200~GeV \cite{phobos-all} and by CMS 
experiment in PbPb collisions at LHC at $\sqrt{s_{NN}}=2.76$~TeV 
 \cite{cms276c}, respectively. 
 As it is noted above, this dependence cannot be reproduced by the  
wounded nucleon model where a  number-of-nucleon-participant scaling is expected. 

Within the above-discussed model of constituent quarks and Landau hydrodynamics, we consider this 
dependence in terms of centrality. The centrality is considered to characterize the degree of 
 overlapping of the volumes of the two colliding nuclei, determined by the 
 impact parameter. 
  The centrality is closely related to the number of nucleon 
participants determined using a Monte Carlo Glauber calculations so that the largest number of 
participants contribute to the most central heavy-ion collisions. Hence the centrality is related 
to the energy released in the collisions, {\sl i.e.} the effective energy, $\varepsilon_{NN}$, which, in the 
framework of the proposed approach, can be defined as a fraction of the c.m. energy available in 
a collision according to the centrality, $\alpha$: 
\begin{equation}
\varepsilon_{NN} = \sqrt{s_{NN}}(1 - \alpha).
\label{Eeff}
\end{equation}
 Conventionally, the data are divided into classes of centrality, or centrality intervals, so 
that $\alpha$ is the average centrality for the centrality interval, {\sl e.g.}\  $\alpha = 0.025$ for 
 $0-5\%$ centrality, which refers to the 5\% most central collisions. 

 Then, for the effective c.m.  energy $\varepsilon_{NN}$, 
Eq.~(\ref{eqn3}) reads 
\begin{eqnarray}
 \nonumber
 &\rho(0)&= 
 \rho_{pp}(0)\, \frac{2N_{\rm ch}}{N_{\rm{part}}\, N_{\rm{ch}}^{pp}} 
\, \sqrt{1 - 
\frac{2 \ln 3}{\ln (2m_p/\varepsilon_{NN})}}\, ,
\\
& & \varepsilon_{NN}=\sqrt{s_{pp}}/3\,,
\label{eqn4}
\end{eqnarray} 
where $N_{\rm{ch}}$ is the  mean multiplicity in central 
nucleus-nucleus collisions measured at  
$\sqrt{s_{NN}}=\varepsilon_{NN}$. The rapidity density 
$\rho_{pp}(0)$ and the multiplicity $N_{\rm{ch}}^{pp}$ 
are taken from the existing data or, where not available, 
calculated using the corresponding  experimental c.m. energy 
fits,\footnote {The E735 power-law
 fit $N_{\rm{ch}}^{pp} = 3.102\,s_{pp}^{0.178}$ 
\cite{pprev} is used, while the linear-log fit
$\rho_{pp}=-0.308+0.276\, \ln(s_{pp})$ 
\cite{pprev} and the power-law fit by CMS \cite{cms276c},
$\rho_{pp}=-0.402+s_{pp}^{0.101}$, are used for
$\sqrt{s_{pp}}\leq$~53~GeV and for $\sqrt{s_{pp}}>$~53~GeV, respectively.} and, 
according to the consideration, the calculations are made at 
$\sqrt{s_{pp}}=3\, \varepsilon_{NN}$. The $N_{\rm ch}$ values are as
well taken from the measurements in central heavy-ion collisions 
wherever available, while for the non-existing data 
the ``hybrid" fit \cite{aditya} combining the linear-logarithmic
 and power-law regularities  is used. This fit is inspired by
 the measurements  as well as by theoretical considerations. 
It is observed that the logarithmic fit well describes
the heavy-ion multiplicity data up to the top RHIC energy
\cite{edward,atlas276c}; however, as the 
collision energy increases above 1$-$2 TeV at the LHC, 
the data clearly show a preference for the 
power-law behaviour \cite{cms276c,atlas276c,alice276r} in
the multiplicity dependence on $\sqrt{s_{NN}}$. From the theoretical 
description point of view, such a c.m. energy dependence is 
expected~\cite{wolschin} as soon as
the logarithmic dependence is considered to characterize the 
fragmentation source(s) while the power-law behaviour is believed to
 come from the gluon-gluon interactions.

In the framework of the model 
 proposed,
 we calculate the 
centrality dependence of the charged particle midrapidity density 
using Eq.~(\ref{eqn4}) to reproduce the centrality data 
shown in Fig.~\ref{Fig1}. 
 One can see that within
 this approach where the collisions are  
derived by the centrality-defined effective c.m. 
energy $\varepsilon_{NN}$, the calculations are in very good 
overall agreement with the measurements independent
 of the collision energy. Similar results are 
obtained as the $N_{\rm part}$-dependence
 of the PHENIX \cite{phenix-all}, STAR \cite{star-all}, 
or CuCu PHOBOS \cite{phobos-all} measurements  
from RHIC and ALICE \cite{alice276c} or ATLAS 
\cite{atlas276c} data from LHC are used (not shown). 
 Some slightly lower values 
 seen in 
 the calculations compared to the data for some 
low-$N_{\rm part}$, {\sl i.e.} for the most peripheral 
 collisions, at $\sqrt{s_{NN}}=19.6$~GeV,  
 look to be due to 
the experimental limitations and the extrapolation
 used in the reconstruction  for the 
measurements in this region of very low multiplicity
 \cite{phobos-all}. 
 The low values 
 obtained within the approach for a few  most central 
 collisions at the LHC energy  can be explained by 
no data on  $N_{\rm ch}^{pp}$ being available  
  at $\sqrt{s_{pp}}>1.8$~TeV.  

Given the obtained agreement between data and the 
calculations and considering the similarity put forward for 
 $\varepsilon_{NN}$ and $\sqrt{s_{NN}}$, one would expect 
the measured centrality data at $\varepsilon_{NN}$
 to follow the  $\sqrt{s_{NN}}$ dependence of the midrapidity 
density in the most central nuclear collisions.
 In Fig.~\ref{Fig2}, the measurements of the 
charged particle pseudorapidity density at midrapidity
 in head-on nuclear collisions are plotted
against the $\sqrt{s_{NN}}$ from a few GeV at
 GSI to a few TeV at the LHC along with the centrality data, 
shown as a function of $\varepsilon_{NN}$, 
 from low-energy RHIC data by STAR at 9.2 GeV 
\cite{star9.2}, and the measurements, shown in 
Fig.~\ref{Fig1}, by PHOBOS  \cite{phobos-all} and 
CMS \cite{cms276c} 
experiments as a function of $\varepsilon_{NN}$. 
The centrality data effective-energy dependence follow well 
the data on the most central collision c.m. energy behaviour. 

We fit the  weighted combination of the 
midrapidity density from 
the head-on collisions 
by the hybrid fit function
\begin{eqnarray}
\nonumber
\lefteqn{\rho(0) =  (-0.306\pm 0.027)
 +   (0.364\pm 0.009)\,
\ln(s_{NN})} \\ 
  & &+(0.0011\pm 
0.0011) \,  s_{NN}^{(0.50\pm 0.06)},
\label{hybrho} 
\end{eqnarray}
which is, as it is noticed above, inspired by the measurements and supported by theoretical 
consideration. The fit combines the linear-log dependence on $\sqrt{s_{NN}}$  observed up to the top RHIC 
energy \cite{phobos-all,phenix-all} and the power-law dependence obtained with the LHC data 
\cite{cms276c,atlas276c,alice276r}. This fit is shown in Fig.~\ref{Fig2} by the dashed line. One can 
see that the fit is as well close to the centrality data.  To clarify, the weighted combination 
of the centrality data are also fitted by the hybrid function, 

\begin{eqnarray}
\nonumber
\lefteqn{\rho(0) =  (0.002\pm 0.080)+ 
(0.646\pm 0.022)\, \ln(\varepsilon_{NN})} \\ 
& & +(0.0003\pm 0.0001) \,  
\varepsilon_{NN}^{(1.158\pm 0.034)},
\label{hybrhoc} 
\end{eqnarray}
where, in addition to the low-energy STAR data and the measurements, 
shown in Fig.~\ref{Fig1}, by the PHOBOS and CMS experiments, 
the midrapidity-density data on the  centrality dependence from 
ALICE \cite{alice276c}, ATLAS \cite{atlas276c}, PHENIX 
\cite{phenix-all} and STAR \cite{star-all} are 
included (not shown). The fit 
 is very close to the fit made to the 
head-on data. From this one can conclude that
 the picture proposed well reproduces the data 
under the assumption of the effective energy deriving
 the multiparticle production process 
pointing to the similarity in all the data from peripheral
 to the most central measurements to 
follow the same energy behaviour. From the fit, we estimate 
the midrapidity-density value to be 
of about 11.5  within 10\% uncertainty 
 in the most central collisions at $\sqrt{s_{NN}}=5.13$~TeV shown 
 in Fig.~\ref{Fig2}. 

In addition to the hybrid fits, in Fig.~\ref{Fig2} 
we show the linear-log fit  \cite{edward} up to 
 the top RHIC energy 
 and the 
 power law fit for the entire energy range 
 to  the most central  collision data
 along with the power law fit to the 
 centrality 
 data  
 weighted. 
 One can see that 
the power-law fit describes well the head-on collision
 measurements (see also \cite{advrev}) and, 
within the errors, does not differ from the linear-log 
or the hybrid functions up to the RHIC 
energies. However, it deviates from the  most central collision  
hybrid fit as soon as the LHC 
measurements are included. The power-law fit to 
the centrality data are much closer to the hybrid fits, 
and it is almost indistinguishable from 
the hybrid fit to the central data up to the head-on
 collision LHC points. Both the power-law 
fits, to the head-on collision  data and to the centrality data, 
 give predictions close to each 
other but lower than the hybrid fits up to some higher c.m. 
energies. Interestingly, using the approach of the effective-energy 
dissipation, one can clearly see the 
transition to a possibly new regime in the multihadron production 
in heavy-ion collisions demonstrated by the data as $\sqrt{s_{NN}}$ 
 increases up to about 600--700 GeV per nucleon. 
 The change in the $\sqrt{s_{NN}}$-dependence from the 
logarithmic to the power-law one seems to be a reason 
of lower-value predictions by theoretical 
models \cite{alice276r}. The change also restrains predictions
 for heavy-ions within the 
universality picture \cite{edward}, which, 
however, gives the correct predictions for ${pp/\bar p}p$ 
\cite{cms-conf}, where both the logarithmic \cite{cmspp7pTn} 
and the power-law \cite{cms276c} functions 
 provide equally good fits to the data up to $\sqrt{s_{pp}}=7$~TeV.  
   
Now, using the effective c.m. energy approach, 
we apply the obtained hybrid function fit of the 
midrapidity density measured in {\it head-on}  
collision data, Eq.~(\ref{hybrho}), to the {\it 
centrality} data, shown in Fig.~\ref{Fig1} as a 
function of $N_{\rm part}$. The calculations are 
shown by the solid lines. One can see that the 
approach well describes the measurements and 
actually follows the predictions by Eq.~(\ref{eqn4}),
 except the LHC data, where it is better 
than the calculations of Eq.~(\ref{eqn4}), 
though slightly overshoots the measurements. 
Similar to the consideration combining constituent 
quarks and Landau hydrodynamics, 
 the low values obtained
using the effective-energy calculations 
 for the very peripheral points at 
 $\sqrt{s_{NN}}=19.6$~GeV
 seem 
 to be due to the  
 difficulties in the measurements.
 A slight 
overestimation of the LHC data is due to the fact that
 the fit (Eq.~(\ref{hybrho})) uses the 
 highest (0-2\% centrality) ATLAS  point of the head-on collisions. 

 In Fig.~\ref{Fig1}, 
 the predictions for
 the $N_{\rm part}$-dependence of the 
 midrapidity density
 are 
 made 
for the 
 forthcoming heavy-ion collisions at $\sqrt{s_{NN}}=5.13$~TeV. 
 Here the 
 centrality and $N_{\rm part}$ values are taken as in
 the 2.76~TeV data.
 The expectations 
show increase of the $\rho(0)$ with $N_{\rm part}$ 
(decrease with centrality) from about 5 to 12. 
The increase  looks to be faster than at $\sqrt{s_{NN}}=2.76$~TeV, 
especially for the peripheral region, 
similar to the change in the behaviour seen as one moves from
 the RHIC 
 to the LHC 
 data, c.f. 200 GeV data and 2.76 TeV data in Fig.~\ref{Fig1}. 
We find that the predictions 
are well reproduced when the LHC data are scaled by a factor
1.43.

Interestingly, within the picture of the effective-energy 
dissipation of constituent quark 
participants, one can explain the observed similarity of 
the midrapidity densities measured in 
${pp/\bar p}p$ interactions and in heavy-ion collisions
 at the same c.m. energy
\cite{indpart,nouicer}
  as well as
 the scaling with the number
 of participants of the midrapidity 
pseudorapidity and transverse energy densities obtained for RHIC 
\cite{eremin,ind,indpart,nouicer,phenixEt} and LHC \cite{aditya} data 
 as soon as the data are calculated 
 in the constituent quark 
framework. Note that  this scaling been  observed also
 for most peripheral collisions may be 
understood in the framework of the approach proposed
 here by considering the most peripheral 
collisions to be driven by nucleon-nucleon interactions.
\\

\begin{figure*}
\centering
\includegraphics[width=10cm]{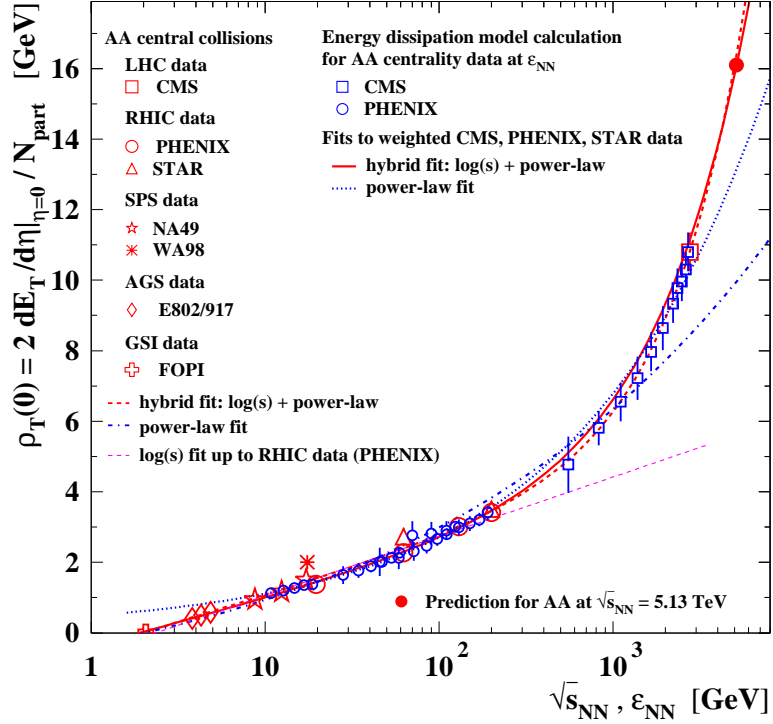}
\caption{The charged particle transverse energy pseudorapidity density
 per participant pair at 
midrapidity as a function of c.m. energy per nucleon, 
 $\sqrt{s_{NN}}$, in central nucleus-nucleus (AA) 
collisions (shown by large symbols), and as a function
 of effective energy, $\varepsilon_{NN}$
(Eq.~(\ref{Eeff})), for AA collisions at different centrality
 (small symbols). The data of central AA collisions  are 
from: the PbPb measurements at LHC by CMS \cite{cms276Et} experiment; 
the AuAu measurements at RHIC by PHENIX \cite{phenix-all,phenixEt} 
and STAR \cite{star200EtRS62.4} 
experiments; the values recalculated in \cite{phenix-all} 
from the measurements at CERN SPS by 
CERES/NA45 \cite{na49} and WA98 \cite{wa98Et} experiments, 
at Fermilab AGS by E802 and E917 
experiments \cite{agsEt}, and at GSI by FOPI Collab. \cite{fopi}.
 The centrality data represent the 
measurements by CMS at the LHC \cite{cms276Et} and 
by PHENIX at RHIC \cite{phenix-all,phenixEt}; the 
CMS and PHENIX data are those from Fig.~\ref{Fig4}, 
while for clarity, just every second point  
of the PHENIX measurements is shown. 
The dashed-dotted line and the dashed line show the fits to 
the central collision data: the power-law fit,
 $\rho_T(0)= -2.29 +1.97 s_{NN}^{0.107}$, and 
the hybrid fit, $\rho_T(0)=-0.447+ 0.327\ln(s_{NN}) +0.002\, s_{NN}^{0.5}$. 
The thin dashed line 
shows the linear-log PHENIX fit  \cite{phenix-all} to
 the central collision data up to the top RHIC 
energy. The dotted line and the solid line show the 
fits to the centrality data: the power-law 
fit, $\rho_T(0)=0.09+ 0.40\,\varepsilon_{NN}^{0.40}$, 
and the hybrid fit, 
$\rho_T(0)=-0.387+ 0.574\ln(\varepsilon_{NN})+ 0.011\, \varepsilon_{NN}^{0.818}$, 
respectively. The fitted centrality 
data include, except of the shown data,  also the measurements 
by STAR \cite{star200EtRS62.4} at 
RHIC (not shown). The solid circle shows the prediction 
for $\sqrt{s_{NN}}=5.13$~TeV.}
\label{Fig3}       
\end{figure*}

\begin{figure*}
\centering
\includegraphics[width=10cm]{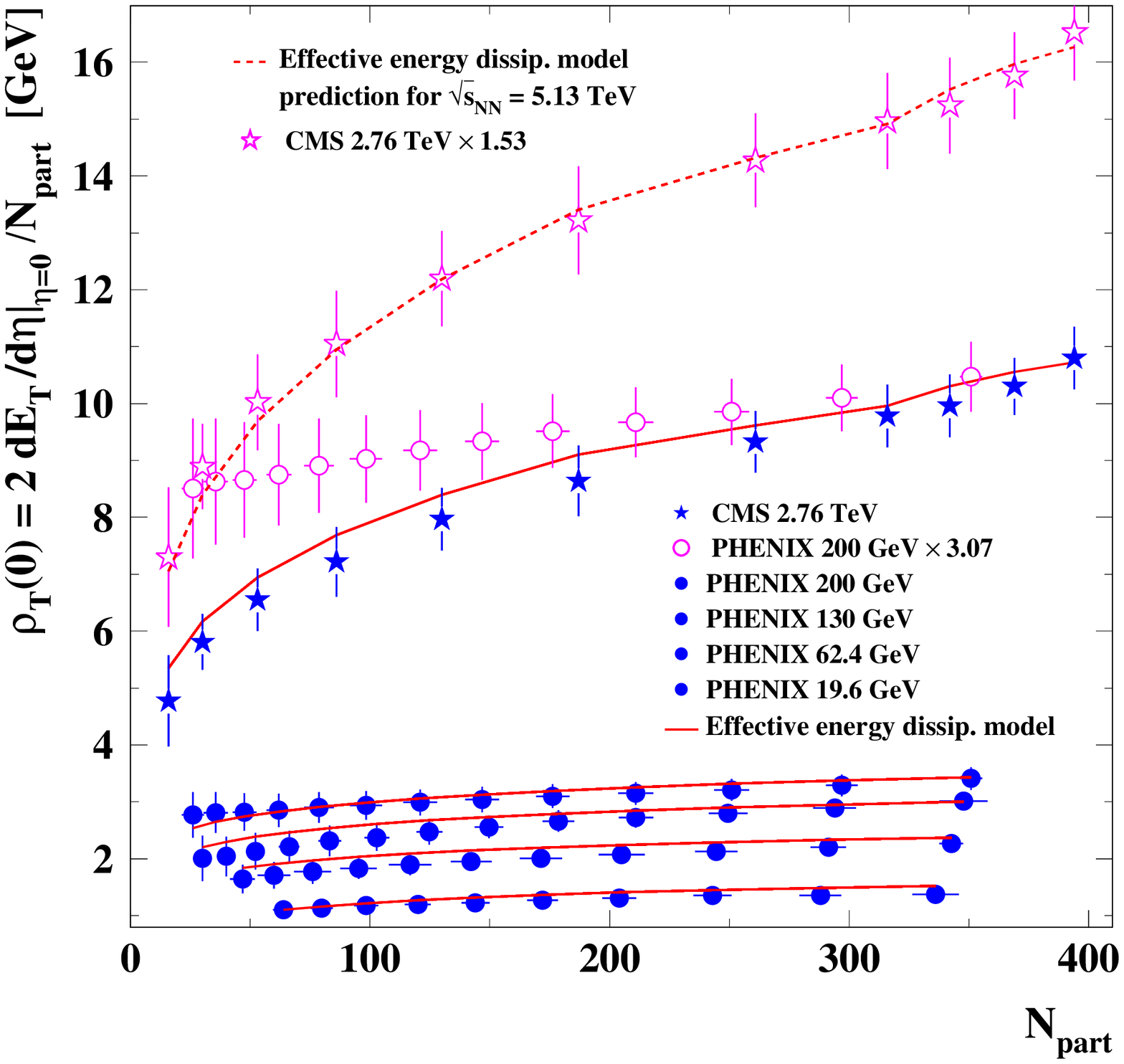}
\caption{The charged particle transverse energy pseudorapidity density 
at midrapidity per participant pair as a function of the number of
participants, $N_{\rm part}$. The solid symbols 
show the data from AuAu collisions at RHIC (circles) by
 PHENIX experiment at $\sqrt{s_{NN}}=19.6$
\cite{phenix-all} and 62.4, 130 and 200~GeV \cite{phenixEt}
 (bottom to top) and  from PbPb collisions 
at LHC by CMS at $\sqrt{s_{NN}}=2.76$~TeV  \cite{cms276Et} (stars).
 The lines show the predictions by the 
effective-energy approach using the hybrid fit to the
 c.m. energy dependence of the midrapidity 
transverse energy density in central heavy-ion collisions 
shown in Fig.~\ref{Fig3}. The open  
circles show the PHENIX measurements at $\sqrt{s_{NN}}=200$~GeV 
multiplied by 3.07, while the open stars 
show the CMS data multiplied by the factor 1.53.}
\label{Fig4}       
\end{figure*}

\noindent
{\textbf 3.} The effective c.m. energy approach applied to the charged particle 
pseudorapidity 
density at midrapidity can be considered to be applied to another
important variable, such as 
the pseudorapidity density  of the transverse energy, 
$\rho_T(\eta)=(2 /N_{\rm part}) \,dE_T/d\eta$, at midrapidity, 
$\eta \approx$~0. The charged 
particle density and the transverse energy density are 
closely related  and, been  studied 
together, provide important characteristics of the
 underlying dynamics of the multihadron 
production. The transverse energy measurements, 
as well as the pseudorapidity data, have been 
shown to be reasonably well modelled by the 
 constituent quark picture \cite{ind,aditya,phenixEt}. 

In Fig.~\ref{Fig3}, the $\sqrt{s_{NN}}$ dependence of the  
charged particle midrapidity transverse 
energy density in pseudorapidity  is displayed as
 measured in head-on collisions at the 
experiments from a few GeV at GSI to a few TeV 
 at the LHC.
 On top of 
these data the centrality data from the  
PHENIX experiment at RHIC \cite{phenixEt} and the CMS 
experiment at LHC \cite{cms276Et} are added 
 as a function of the effective c.m. energy $\varepsilon_{NN}$.  
 Similarly to the case of 
the charged particle density at midrapidity, the 
$E_T$ density data show the complementarity of
 these two types of measurements.

To better trace the similarity in the energy dependence 
of the central collision and the 
centrality-dependent data, we fit the data by the
 hybrid function, as is done in Fig.~\ref{Fig2} 
for the particle psudorapidity densities. For the 
central collisions one gets

\begin{eqnarray}
\nonumber
 \lefteqn{\rho_T(0) =  (-0.447\pm 0.014)+ (0.327\pm 0.011)\, 
\ln(s_{NN})}\\ 
& & +(0.002\pm 0.003)\,  s_{NN}^{0.50\pm 0.08},
\label{hybet}
\end{eqnarray}

and similar fit to  the centrality data reads
\begin{eqnarray}
\nonumber
\lefteqn{\rho_T(0)  =  (-0.387\pm 0.090)+ (0.574\pm 0.032)\, 
\ln(\varepsilon_{NN})} \\
& & +(0.011\pm 0.005)\,  \varepsilon_{NN}^{0.818\pm 0.064}\,.
\label{hybetc}
\end{eqnarray}
 The fits are shown in Fig.~\ref{Fig3}.
 The data from 
different experiments are weighted, and 
the fit of the effective c.m. energy $\varepsilon_{NN}$
 includes the 
STAR measurements in addition to the PHENIX ones.
 One can see that the two fits are amazingly 
close to each other for the entire energy range
 allowing to  conclude that the effective-energy 
approach provides a good description of the $E_T$ 
production in heavy-ion collisions. We estimate 
the value of $\rho_T(0)$ to be about 16.1~GeV  
with about 10\% uncertainty for most central 
collisions at $\sqrt{s_{NN}}=5.13$~TeV.

As is obtained above for the midrapidity pseudorapidity-density
 energy dependence (Fig.
\ref{Fig2}), in Fig.~\ref{Fig3} the LHC data demonstrate
 a clear departure from the linear-log
regularity  \cite{phenix-all} in the region of $\sqrt{s_{NN}} \simeq 
500-700$~GeV.
 This observation 
supports a possible transition to  a new regime in heavy-ion
collisions
 at $\sqrt{s_{NN}}$ above a few 
hundred GeV as indicated by the midrapidity density in 
Fig.~\ref{Fig2}. In Fig.~\ref{Fig3}, we 
also show the power-law fits  to the central collision
 measurements 
 and 
 to the centrality data weighted.
 One can see that the power-law fit to the central 
 collision data underestimates the LHC 
measurement at 2.76 TeV data and deviates
 from the data at $\sqrt{s_{NN}}\sim 1$~TeV. 
However, the power-law fit to the centrality describes well the data
 in the full available c.m.-energy region,
though lies slightly lower than the hybrid
  fit, Eq.~(\ref{hybetc}). Meantime, this fit 
overestimates the data below $\sqrt{s_{NN}} \approx 10$~GeV, 
similarly to the case of the multiplicity 
data on centrality, Fig.~\ref{Fig2}. Interestingly, 
the shown  power-law fit curve to the 
centrality data is similar to that obtained by 
CMS for $\sqrt{s_{NN}}\ge 8.7$~GeV \cite{cms276Et}; moreover, 
fitting all the $E_T$ {\it centrality} data \`a-la CMS, 
one finds a good fit to the data by $
\rho_T(0)=0.43 \,\varepsilon_{NN}^{0.20}$ (not shown)
 which resembles the CMS fit, $\rho_T(0)=0.46 \,
\sqrt{s_{NN}}^{0.20}$, to the {\it head-on} collision data. 
This again demonstrates  the multihadron 
production in heavy-ion collisions to be well
 described by the effective c.m. energy dissipation  
 picture.

To further exploit the effective-energy approach 
with the centrality data, in Fig.~\ref{Fig4} we 
show the $N_{\rm part}$ dependence of the centrality 
data from Fig.~\ref{Fig3} along with the 
central collision data fit, Eq.~(\ref{hybet}), but as a 
function of the centrality-dependent 
c.m. effective energy $\varepsilon_{NN}$. One can 
see that the fit well describes the data.
 Interestingly, 
 the RHIC
 data at $\sqrt{s_{NN}}=200$~GeV 
 shown 
 scaled by 3.07
demonstrate much less decrease as the centrality 
increases (more peripheral data), than that observed 
for the LHC data. This is different for the 
pseudorapidity density of charged particles at 
midrapidity measurements (see Fig.~\ref{Fig1}). In 
contrast to the scaled RHIC data, the effective-energy approach follows 
well the LHC 
 measurements. 

 In  Fig.~\ref{Fig4}
 we show
 the 
 predictions made  within the
 effective-energy dissipation approach
 for the $N_{\rm part}$-dependence of the
midrapidity transverse energy density for the future 
heavy-ion 
 collisions at $\sqrt{s_{NN}}=5.13$~TeV. 
 The predictions show more rapid increase of the 
 $\rho_T(0)$ with $N_{\rm part}$ (decrease with centrality) 
than at $\sqrt{s_{NN}}=2.76$~TeV, especially 
for the peripheral region, similar to the change
 in the behaviour seen as one moves from the RHIC 
measurements to the LHC data and similar to that 
obtained for the midrapidity density, 
Fig.~\ref{Fig1}. We find that the predictions made
 are well reproduced as the LHC data are scaled 
by a numerical factor 1.53.

The similarity 
between the features of the midrapidity density
and the midrapidity transverse-energy density observed 
here (and in
\cite{ind,aditya,phenixEt}) in the constituent quark framework and the
successful applicability of the effective-energy approach to both the
variables show that, using the main assertions of the Landau approach, one
is able to correctly estimate fractions of the energy dissipated into
produced particles, despite the Landau model is a 1$+$1 model and 
therefore
does not take into account the transverse expansion of the system. This
seems to reflect the fact that the inclusion of the transverse expansion
in the Landau model does not change the scaling of the observables under
study \cite{LandaupT,Feinberg}.
\\

\noindent
{\textbf 4.} In summary, we analyzed the midrapidity pseudorapidity 
density 
of charged particles 
and of the transverse energy measured in 
nucleus-nucleus collisions in the whole available range 
of the collision c.m. energy per nucleon, $\sqrt{s_{NN}}$, 
 from a few GeV 
 up to a few TeV. 
The dependencies of these key variables
 on the c.m. energy per nucleon and on the number of 
participants (or centrality) have been revealed within 
the approach of the dissipation of the
effective energy pumped in by the participants of the
 collisions, which forms the effective-energy budget in the multiparticle 
production process. 
Namely, the model of constituent quarks 
combined with Landau hydrodynamics is applied to 
reproduce the midrapidity-density dependence on 
the number of participants. This approach, proposed
 earlier in \cite{edward} and pointed to the 
universality of the multihadron production in different
 types of collisions up to the top RHIC  
energy allows one to well predict the LHC measurements 
in ${pp/\bar p}p$ interactions on the midrapidity 
density of charged particles. Within this picture, 
we find that the dependence of the 
pseudorpaidity density at midrapidity from the 
RHIC to LHC data is well reproduced as soon as the 
effective c.m. energy variable is introduced as the
 centrality-defined fraction of the collision 
c.m. energy. Based on this finding, it is shown that
 the most central collision data and the 
centrality-dependent data follow a similar $\sqrt{s_{NN}}$
 dependence obtained for the central collision 
data as soon as the centrality data is rescaled to
 the effective energy. The hybrid fit,
combining the linear-log and the power-law c.m. energy dependencies of the 
head-on collision 
  data, 
 is found to well reproduce the 
 dependence of the midrapidity densities on the
 number of participants within the effective-energy 
 approach. 
 Similar observations are made for the
 transverse energy midrapidity-density 
measurements.
 For both the variables studied, a 
clear departure of the data as a function of the 
effective c.m. energy from the linear-log 
dependence to the power-law one is observed 
at $\sqrt{s_{NN}} \simeq 500-700$~GeV indicating a possible 
transition to a new regime in heavy-ion collisions. 
The data at $\sqrt{s_{NN}}\sim 1$~TeV would be 
extremely useful to clarify the observations made here. 
Based on the hybrid fits in the framework 
of the discussed approach, the predictions
 for the energy and the number-of-participant 
dependencies for the measurements in the forthcoming 
heavy-ion runs at LHC at $\sqrt{s_{NN}}$ above 
 5~TeV are made.

\vspace{0.5cm}
\noindent
{\textbf Acknowledgement} \\
The authors are grateful to the International Symposium on Multiparticle Dynamics (ISMD-2014) and Workshop on Particle Correlations and Femptoscopy (WPCF-2014) organizers for kind invitation, warm hospitality and financial supports.
\vspace{0.5cm}
\noindent


\begin{thebibliography}{}
%
%
\bibitem{Heisenberg} W. Heisenberg, {\rm Zs. Phis.} \textbf {126}, 569 
(1949)
\bibitem{Fermi} E. Fermi, {\rm Prog. Theor. Phys.}\textbf{5}, 570 (1950) 
\bibitem{Landau} L. D. Landau, {\rm Izv. Akad. Nauk}: Ser. Fiz. 
\textbf{17}, 51 (1953). English translation: {\em  Collected Papers of 
L.D. Landau}, ed. by D. Ter-Haarp (Pergamon, Oxford, 1965), p. 569. Reprinted in: {\em Quark-Gluon Plasma: Theoretical Foundations}, ed. by J. Kapusta, B. M\"uller, J. Rafelski (Elsevier, Amsterdam, 2003), p. 283
\bibitem{wound} A. Bia{\l}as, B. Bleszy\'nski, W. Czy\.z, {\rm Nucl. Phys. 
B }\textbf{111}, 461 (1976); for review, see: A. Bia{\l}as, {\rm J. Phys. 
G } \textbf {35}, 044053 (2008)
\bibitem{eremin} S.~Eremin, S.~Voloshin, {\rm Phys. Rev. C } \textbf{67}, 
064905 (2003)
\bibitem{ind} P.K. Netrakanti, B. Mohanty,  {\rm Phys. Rev. C } 
\textbf{70}, 027901 (2004)
\bibitem{indpart} B. De, S. Bhattacharyya, {\rm Phys. Rev. C } 
\textbf{71}, 024903 (2005)
\bibitem{nouicer} R. Nouicer,  {\rm AIP Conf. Proc.} \textbf{ 828} (2006) 
11, \textbf{842}, 86 (2006), {\rm Eur. Phys. J. C } \textbf {49}, 281 
(2007)
\bibitem{phobos-sim}  B.B.~Back et al., 
nucl-ex/0301017, {\rm Phys. Rev. C } \textbf{74}, 021902 (2006)
\bibitem{book}  W. Kittel, E.A. De Wolf, {\it Soft Multihadron Dynamics} 
(World Scientific, Singapore, 2005)
\bibitem{advrev} R. Singh, L. Kumar, P.K. Netrakanti, B. Mohanty, {\rm 
Adv. High Energy Phys.} \textbf{2013}, 761474 (2013)
\bibitem{edward} E.K.G. Sarkisyan, A.S. Sakharov, {\rm Eur. Phys. J. C
  } \textbf{ 70}, 533 (2010), {\rm AIP Conf. Proc. } \textbf{828}, 35
  (2006)
\bibitem{constitq} For review, see: V.V. Anisovich, N.M. Kobrinsky,  J. Nyiri, Yu.M. Shabelsky, {\it Quark Model and High Energy Collisions}
(World Scientific, Singapore, 2004)
\bibitem{cms-conf}  R. Rougny (for the CMS Collab.), {\rm Nucl. Phys.  B 
(Proc. Suppl.) } \textbf{207-208}, 29 (2010) 
\bibitem{edward14} A.N. Mishra, R. Sahoo, E.K.G. Sarkisyan, A.S. Sakharov, 
{\rm Eur. Phys. J. C } \textbf{74}, 3147 (2014) 
\bibitem{phobos-all} B. Alver et al., {\rm Phys. Rev. C } \textbf{83}, 
024913 (2011).
\bibitem{cms276c} CMS Collab., S. Chatrchyan et al.,  {\rm J. High Energy 
Phys.} \textbf{08}, 141 (2011)
 \bibitem{pprev} J.F. Grosse-Oetringhaus, K. Reygers,  {\rm J. Phys. G } 
 \textbf {37}, 083001 (2010)
\bibitem{atlas276c} ATLAS Collab., G. Aad et al.,  {\rm Phys. Lett. B } 
\textbf{710}, 363 (2012)
\bibitem{alice276r}  ALICE Collab., K. Aamodt et al., {\rm  Phys. Rev. 
Lett. } \textbf{105}, 252301 (2010)
\bibitem{brahms} BRAHMS Collab., I.G. Bearden et al., {\rm Phys. Rev. 
Lett. 
} \textbf{94}, 162301 (2005)
\bibitem{phenix-all} PHENIX Collab., S.S. Adler et al., {\rm Phys. Rev. C 
} 
\textbf{74}, 049901 (2005), ibid. \textbf {71} (2005)
  034908 (E); A. Milov (for the PHENIX Collab.), {\rm  J. Phys. Conf. 
Ser.} \textbf{5}, 17 (2005)
\bibitem{star-all} STAR Collab., B.I. Abelev et al., {\rm Phys. Rev. C } 
\textbf{79}, 034909 (2009) 
\bibitem{star9.2}  STAR Collab., B.I. Abelev et al., {\rm Phys. Rev. C } 
\textbf{81}, 024911 (2010)
\bibitem{na45} F. Ceretto (for the CERES/NA45 Collab.), {\rm Nucl. Phys. A 
} \textbf{638}, 467c (1998)
\bibitem{na49} F. Sikl\'er (for the NA49 Collab.), {\rm Nucl. Phys. A } 
\textbf{661}, 45c (1998); 
NA49 Collab., S.V. Afanasiev et al., {\rm Phys. Rev. C } \textbf{66}, 
054902 (2002)
\bibitem{ags} E802 Collab., L. Ahle et al., {\rm Phys. Rev. C } 
\textbf{59}, 2173 (1999);
E917 Collab., B.B. Back et al., {\rm Phys. Rev. Lett. } \textbf{86}, 1970 
(2001)
\bibitem{fopi} Estimated in \cite{phenix-all} from: FOPI Collab., W. 
Reisdorf et al., {\rm Nucl. Phys. A} \textbf{612}, 493 (1997)
\bibitem{alice276c} ALICE Collab., K. Aamodt et al.,  {\rm Phys. Rev. 
Lett. 
} \textbf{106}, 032301 (2011)
\bibitem{aditya} R. Sahoo, A. N. Mishra, {\rm Int.  J. Mod. Phys. E } 
\textbf {23}, 1450024  (2014)
\bibitem{wolschin} G. Wolschin, {\rm J. Phys. G } \textbf{40}, 045104 
(2013)
\bibitem{cmspp7pTn} CMS Collab., V. Khachatryan et al., {\rm Phys. Rev. 
Lett. } \textbf{105}, 022002 (2010)
\bibitem{phenixEt} PHENIX Collab., S.S. Adler et al., {\rm Phys. Rev. C } 
\textbf{89}, 044905 (2014)
\bibitem{cms276Et} CMS Collab., S. Chatrchyan et al., {\rm Phys. Rev. 
Lett. 
} \textbf{109}, 152303 (2012)
\bibitem{star200EtRS62.4} J. Adams et al., {\rm Phys. Rev. C } 
\textbf{70}, 
054907 (2004), R. Sahoo, PhD Thesis (Utkal University, 
2007), arXiv:0804.1800
\bibitem{wa98Et} WA98 Collab., M.M. Aggarval et al., {\rm Eur. Phys. J. C 
} \textbf{18}, 651 (2001)
\bibitem{agsEt} E802 Collab., T. Abbott et al., {\rm Phys. Rev. C } 
\textbf{63}, 064602 (2001)
\bibitem{LandaupT} G.A. Milekhin, Sov. Phys. JETP \textbf{35}, 829 (1959); 
E.V. Shuryak, Sov. J. Nucl. Phys. \textbf{20}, 295 (1974)
 \bibitem{Feinberg} For review, see: E.L. Feinberg, in {\it Relativistic
Heavy Ion Physics} (ed. by L.P. Csernai, D.D. Strottman): Int. Rev. Nucl.
Phys., vol. 6 (World Scientific, Singapore, 1991), p 341


\end{thebibliography}
\end{document}